%% file: arXiv.tex
\documentclass[runningheads,a4paper]{llncs}

\usepackage{graphicx}
\usepackage[space]{grffile}
\usepackage{latexsym}
\usepackage{amsfonts,amsmath,amssymb}
\usepackage{url}
\usepackage[utf8]{inputenc}
\usepackage{hyperref}
\hypersetup{colorlinks=false,pdfborder={0 0 0}}
\usepackage{textcomp}
\usepackage{longtable}
\usepackage{multirow,booktabs}

\usepackage{amssymb}
\newcommand{\ceiling}[1]{\left\lceil{#1}\right\rceil}

\newtheorem{Property}{Property}
\newtheorem{Lemma}{Lemma}
\newtheorem{Corollary}{Corollary}

 \def\myendproof{{\ \vbox{\hrule\hbox{%
   \vrule height1.3ex\hskip0.8ex\vrule}\hrule }}\par}
 \renewenvironment{proof}{\noindent{\bf Proof. }}{\myendproof}  
  
\newif\ifpaper

\paperfalse

\usepackage{tikz}
\usetikzlibrary{arrows}
\usepackage{sansmath}
\tikzset{>=latex}
\usepackage{enumerate}

\begin{document}

\title{A Note on Modeling Self-Suspending Time as Blocking Time in
  Real-Time Systems\thanks{This paper is supported by DFG, as part of
    the Collaborative Research Center SFB876
    (http://sfb876.tu-dortmund.de/). This work was also partially supported by National Funds through FCT/MEC (Portuguese Foundation for Science and Technology) and co-financed by ERDF (European Regional Development Fund) under the PT2020 Partnership, within project UID/CEC/04234/2013 (CISTER); also by FCT/MEC and the EU ARTEMIS JU within project(s) ARTEMIS/0003/2012 - JU grant nr. 333053 (CONCERTO) and ARTEMIS/0001/2013 - JU grant nr. 621429 (EMC2).}}
\titlerunning{A Note on Modeling Suspension as Blocking}
\author{Jian-Jia Chen\inst{1}, Wen-Hung Huang\inst{1}, and Geoffrey Nelissen\inst{2}}
\institute{TU Dortmund University, Germany\\
Email: jian-jia.chen@tu-dortmund.de, wen-hung.huang@tu-dortmund.de
\and
CISTER/INESC-TEC, ISEP, Polytechnic Institute of Porto, Portugal \\
Email: grrpn@isep.ipp.pt
}

\maketitle

\begin{abstract}
  This report presents a proof to support the correctness of the schedulability test for self-suspending real-time task systems proposed by Jane W. S. Liu in her book titled ``Real-Time Systems'' (Pages 164-165). The same concept was also implicitly used by Rajkumar, Sha, and Lehoczky in RTSS 1988 (Page 267) for analyzing self-suspending behaviour due to synchronization protocols in multiprocessor systems.
\end{abstract}
\section{Introduction}

This report presents a proof to support the correctness of the
schedulability test for self-suspending real-time task systems proposed by Jane
W. S. Liu in her book titled "Real-Time Systems" \cite[Pages 164-165]{Liu:2000:RS:518501}.
The same concept was also implicitly used by Rajkumar, Sha, and Lehoczky \cite[Page 267]{DBLP:conf/rtss/RajkumarSL88} for analyzing self-suspending behaviour due to synchronization protocols in multiprocessor systems.

The system model and terminologies are defined as follows: We assume a system composed of $n$ sporadic self-suspending tasks. A sporadic task $\tau_i$ is released repeatedly, with each such invocation called a
job. The $j^{th}$ job of $\tau_i$, denoted $\tau_{i,j}$, is released
at time $r_{i,j}$ and has an absolute deadline at time $d_{i,j}$. Each
job of any task $\tau_i$ is assumed to have a worst-case execution time $C_i$. Each job of task $\tau_i$ 
suspends for at most $S_i$ time units (across all of its suspension phases). When a job suspends itself, 
the processor can execute another job. 
The response time
of a job is defined as its finishing time minus its release
time. Successive jobs of the same task are required to execute in
sequence. Associated with each task $\tau_i$ are a period (or minimum inter-arrival time) $T_i$, which
specifies the minimum time between two consecutive job releases of
$\tau_i$, and a relative deadline $D_i$, which specifies the maximum amount of time a job can take to complete its execution after its release, i.e., $d_{i,j}=r_{i,j}+D_i$. The worst-case response
time $R_i$ of a task $\tau_i$ is the maximum response time among all its
jobs.  The utilization of a task $\tau_i$ is defined as $U_i=C_i/T_i$.

In this report, we focus on constrained-deadline task systems, in which $D_i \leq T_i$ for every task $\tau_i$. We only consider preemptive fixed-priority scheduling on a single processor, in which each task is assigned with a unique priority level. We assume that the priority assignment is given.

%We will focus on the analysis of task $\tau_k$. There are $k-1$ higher-priority tasks, i.e., $\tau_1, \tau_2, \ldots, \tau_{k-1}$, than task $\tau_k$. 
We assume that the tasks are numbered in a decreasing priority order. That is, a task with a smaller index has higher priority than any task with a higher index, i.e., task $\tau_i$ has a higher-priority level than task $\tau_{i+1}$. When performing the schedulability analysis of a specific task $\tau_k$, we assume that $\tau_1, \tau_2, \ldots, \tau_{k-1}$ are already verified to meet their deadlines, i.e., that $R_i \leq D_i, \forall \tau_i \mid 1 \leq i \leq k-1$. We also classify the $k-1$ higher-priority tasks into two sets: ${\bf T}_1$ and ${\bf T}_2$. A task $\tau_i$ is in ${\bf T}_1$ if $C_i \geq S_i$; otherwise, it is in ${\bf T}_2$. 

\section{Model Suspension Time and Blocking Time}

To analyze the worst-case response time (or the schedulability) of task $\tau_k$, we usually need to quantify the worst-case interference caused by the higher-priority tasks on the execution of any job of task $\tau_k$. In the ordinary sequential sporadic real-time task model, i.e., when $S_i=0$ for every task $\tau_i$, the so-called critical instant theorem by Liu and Layland \cite{Liu_1973} is commonly adopted. That is, the worst-case response time of task $\tau_k$ (if it is less than or equal to its period) happens for the first job of task $\tau_k$ when $\tau_k$ and all the higher-priority tasks release a job synchronously and the subsequent jobs are released as early as possible (i.e., with a rate equal to their period). 

However, as proven in \cite{ecrts15nelissen}, this definition of the critical instant does not hold for self-suspending sporadic tasks.  
In \cite[Pages 164-165]{Liu:2000:RS:518501}, Jane W. S. Liu proposed a solution to study the schedulability of self-suspending tasks by modeling the \emph{extra delay} suffered by a task $\tau_k$ due to the self-suspending behavior of the tasks as a blocking time denoted as $B_k$ and defined as follows:
\begin{itemize}
\item The blocking time contributed from task $\tau_k$ is $S_k$.
\item A higher-priority task $\tau_i$ can only block the execution of task $\tau_k$ by at most $b_i=min(C_i, S_i)$ time units.
\end{itemize}
%As a result, if $S_i < C_i$, then the blocking time contributed from task $\tau_i$ is at most $S_i$. Therefore, the contribution of a higher-priority task $\tau_i$  to $B_k$ is at most $b_i=min(C_i, S_i)$.
Therefore, 
\begin{equation}
\label{eq:Bk}
B_k = S_k + \sum_{i=1}^{k-1} b_i.
\end{equation}

In \cite{Liu:2000:RS:518501}, the blocking time is then used to derive a utilization-based schedulability test for rate-monotonic scheduling. Namely, it is stated that if $\frac{C_k+B_k}{T_k} + \sum_{i=1}^{k-1} U_i \leq k (2^{\frac{1}{k}}-1)$, then task $\tau_k$ can be feasibly scheduled by using rate-monotonic scheduling if $T_i=D_i$ for every task $\tau_i$ in the given task set. If the above argument is correct, we can further prove that a constrained-deadline task $\tau_k$ can be feasibly scheduled by the fixed-priority scheduling if
\begin{equation}
\label{eq:TDA-suspension}
\exists t \mid 0 < t \leq D_k, \qquad C_k + B_k + \sum_{i=1}^{k-1}\ceiling{\frac{t}{T_i}} C_i \leq t.
\end{equation}

The same concept was also implicitly used by Rajkumar, Sha, and Lehoczky \cite[Page 267]{DBLP:conf/rtss/RajkumarSL88} for analyzing self-suspending behaviour due to synchronization protocols in multiprocessor systems. To account for the self-suspending behaviour, it reads as follows:\footnote{We rephrased the wordings and notation to be consistent with this report.}
\begin{quote}
\emph{For each higher priority job $J_i$ on the processor that suspends on global semaphores or for other reasons, add the term $min(C_i, S_i)$ to $B_k$, where $S_i$ is the maximum duration that $J_i$ can suspend itself. The sum ... yields $B_k$, which in turn can be used in 
$\frac{C_k+B_k}{T_k} + \sum_{i=1}^{k-1} U_i \leq k (2^{\frac{1}{k}}-1)$ to determine whether the current task allocation to the processor is schedulable.}
\end{quote}

However, as there is no proof in \cite{Liu:2000:RS:518501,DBLP:conf/rtss/RajkumarSL88} to support the correctness of the above tests, we present a proof in the next section of this report.

%%% Local Variables:
%%% mode: latex
%%% TeX-master: "authorea_build/authorea_paper.tex"
%%% End:

\ifpaper
\newtheorem{Property}{Property}
\newtheorem{Lemma}{Lemma}
\newtheorem{Corollary}{Corollary}
\fi
\section{Our Proof}  

This section provides the proof to support the correctness of the test in Eq. \eqref{eq:TDA-suspension}. First, it should be easy to see that we can convert the suspension time of task $\tau_k$ into computation. This has been proven in many previous works, e.g., Lemma~3 in \cite{Liu_2014} and Theorem~2 in \cite{ecrts15nelissen}. Yet, it remains to formally prove that the additional interference due to the self-suspension of a higher-priority task $\tau_i$ is upper-bounded by $b_i=min(C_i, S_i)$. The interference to be at most $C_i$ has been provided in the literature as well, e.g., \cite{Rajkumar_1990,Liu_2014}. However, the argument about blocking task $\tau_k$ due to a higher-priority task $\tau_i$ by at most $S_i$ amount of time is not straightforward. 

From the above discussions, we can greedily convert the suspension time of task $\tau_k$ to its computation time. For the sake of notational brevity, let $C_k'$ be $C_k + S_k$. We call this converted version of task $\tau_k$ as task $\tau_k'$. Our analysis is also based on very simple properties and lemmas enunciated as follows:

\begin{Property}
\label{prop:lower-priority}
In a preemptive fixed-priority schedule, the lower-priority jobs do not impact the schedule of the higher-priority jobs.
\end{Property}

%\begin{lemma}
%\label{lemma:remove-lower-priority}
%In a preemptive fixed-priority schedule, removing a lower-priority job arrived at time $t$ does not affect the schedule of the higher-priority jobs after time $t$.
%\end{lemma}
%\begin{proof}
%This is a direct consequence of Property~\ref{prop:lower-priority}.
%\end{proof}

\begin{Lemma}
\label{lemma:remove-same-task}
In a preemptive fixed-priority schedule, if the worst-case response time of task $\tau_i$ is no more than its period $T_i$, preventing the release of a job of task $\tau_i$ does not affect the schedule of any other job of task $\tau_i$.
\end{Lemma}
\begin{proof}
Since the worst-case response time of task $\tau_i$ is no more than its period, any job $\tau_{i,j}$ of task $\tau_i$ completes its execution before the release of the next job $\tau_{i,j+1}$. Hence, the execution of $\tau_{i,j}$ does not directly interfere with the execution of any other job of $\tau_i$, which then depends only on the schedule of the higher priority jobs. Furthermore, as stated in Property~\ref{prop:lower-priority}, the removal of $\tau_{i,j}$ has no impact on the schedule of the higher-priority jobs, thereby implying that the other jobs of task $\tau_i$ are not affected by the removal of $\tau_{i,j}$.
\end{proof}

We can prove the correctness of Eq. \eqref{eq:TDA-suspension} by using a similar proof than for the critical instant theorem of the ordinary sporadic task model.
Let $R_k'$ be the minimum $t$ greater than $0$ such that  Eq. \eqref{eq:TDA-suspension} holds, i.e., $C'_k + B_k + \sum_{i=1}^{k-1}\ceiling{\frac{R_k'}{T_i}} C_i = R_k'$ and $\forall 0 < t < R_k'$ $C'_k + B_k + \sum_{i=1}^{k-1}\ceiling{\frac{t}{T_i}} C_i > t$. The following theorem shows that $R_k'$ is a safe upper bound if the worst-case response time of task $\tau_k'$ is no more than $T_k$.

\begin{theorem}
\label{theorem:critical}
 $R_k'$ is a safe upper bound on the worst-case response time of task $\tau_k'$ if the worst-case response time of $\tau_k'$ is not larger than $T_k$.
\end{theorem}
\begin{proof}
Let us consider the task set $\tau'$ composed of $\left\{\tau_1, \tau_2, \ldots, \tau_{k-1}, \tau_k', \tau_{k+1}, \ldots \right\}$ and let $\Psi$ be a schedule of $\tau'$ that generates the worst-case response time of $\tau_k'$.  

The proof is built upon the two following steps:
\begin{enumerate}
\item We discard all the jobs that do not contribute to the worst-case response time of $\tau_k'$ in the schedule $\Psi$. We follow an inductive strategy by iteratively inspecting the schedule of the higher priority tasks in $\Psi$, starting with $\tau_{k-1}$ until the highest priority task $\tau_1$. At each iteration, a time instant $t_j$ is identified such that $t_j \leq t_{j+1}$ ($1 \leq j < k$). Then, all the jobs of task $\tau_j$ released before $t_j$ are removed from the schedule and, if needed, replaced by an artificial job mimicking the interference caused by the residual workload of task $\tau_j$ at time $t_j$ on the worst-case response time of $\tau_{k}'$. 
\item The final reduced schedule is analyzed so as to characterize the worst-case response time of $\tau_k'$ in $\Psi$. We then prove that $R_k'$ is indeed an upper bound on the worst-case response time of $\tau_k'$.
\end{enumerate}

\noindent{\bf Step 1: Reducing the schedule $\Psi$} 

During this step, we iteratively build an artificial schedule $\Psi^j$ from $\Psi^{j+1}$ (with $1 \leq j < k$) so that the response time of $\tau_{k}'$ remains identical. At each iteration, we define $t_j$ for task $\tau_j$ in the schedule $\Psi^{j+1}$ (with $j=k-1, k-2, \ldots, 1$) and build $\Psi^j$ by removing all the jobs released by $\tau_j$ before $t_j$.

~

\noindent\textit{Basic step (definition of $\Psi^k$ and $t_k$):} 

Suppose that the job $J_{k}$ of task $\tau_k'$ with the largest response time in $\Psi$ arrives at time $r_k$ and finishes at time $f_k$. We know by Property~\ref{prop:lower-priority} that the lower priority tasks $\tau_{k+1}, \tau_{k+2}, \ldots, \tau_n$ do not impact the response time of $J_{k}$. Moreover, since we assume that the worst-case response time of task $\tau_k'$ is no more than $T_k$, Lemma~\ref{lemma:remove-same-task} proves that removing all the jobs of task $\tau_k'$ but $J_{k}$ has no impact on the schedule of $J_{k}$. Therefore, let $\Psi^k$ be a schedule identical to $\Psi$ but removing all the jobs released by the lower priority tasks $\tau_{k+1}, \ldots, \tau_n$ as well as all the jobs released by $\tau_k'$ at the exception of $J_{k}$. The response time of $J_{k}$ in $\Psi^{k}$ is thus identical to the response time of $J_{k}$ in $\Psi$.

We define $t_k$ as the release time of $J_k$ (i.e., $t_k = r_k$).

~

\noindent\textit{Induction step (definition of $\Psi^j$ and $t_j$ with $1 \leq j < k$):}

Let $r_j$ be the arrival time of the last job released by $\tau_j$ before $t_{j+1}$ in $\Psi^{j+1}$ and let $J_{j}$ denote that job. %There are a two possible cases:
%\begin{itemize}
%\item $J_{j}$ completed its execution no later than $t_{j+1}$. Then, we simply set $t_j$ to $t_{j+1}$ and generate $\Psi^j$ by removing all the jobs of task $\tau_j$ arrived before $t_{j+1}$ in the schedule $\Psi^{j+1}$. By Lemma~\ref{lemma:remove-same-task} and Property \ref{prop:lower-priority}, removing all the jobs of task $\tau_j$ arrived before $t_{j+1}$ has no impact on the schedule of the higher-priority jobs (jobs released by $\tau_1, \ldots, \tau_{j-1}$) and the jobs of $\tau_j$ released after $t_{j+1}$. Moreover, because no task with a priority lower than $\tau_j$ executes jobs before $t_{j+1}$ in $\Psi^{j+1}$, removing the jobs released by $\tau_j$ before $t_{j+1}$ does not impact the schedule of the jobs of $\tau_{j+1}, \ldots, \tau_{k}$. The response time of $J_{k}$ in $\Psi^j$ thus remains unchanged in comparison to its response time in $\Psi^{j+1}$. 
%\item $J_{j}$ did not complete its execution before $t_{j+1}$.
Removing all the jobs of task $\tau_j$ arrived before $r_j$ has no impact on the schedule of any other job released by $\tau_j$ (Lemma~\ref{lemma:remove-same-task}) or any higher priority job released by $\tau_1, \ldots, \tau_{j-1}$ (Property \ref{prop:lower-priority}). Moreover, because by the construction of $\Psi^{j+1}$, no task with a priority lower than $\tau_j$ executes jobs before $t_{j+1}$ in $\Psi^{j+1}$, removing the jobs released by $\tau_j$ before $t_{j+1}$ does not impact the schedule of the jobs of $\tau_{j+1}, \ldots, \tau_{k}$. Therefore, we can safely remove all the jobs of task $\tau_j$ arrived before $r_j$ without impacting the response time of $J_{k}$. Two cases must then be considered:
\begin{enumerate}[(a)]
\item $\tau_j \in {\bf T}_1$, i.e., $S_j < C_j$. In this case, we analyze two different subcases:
\begin{itemize}
\item $J_{j}$ completed its execution before or at $t_{j+1}$. By Lemma~\ref{lemma:remove-same-task} and Property \ref{prop:lower-priority}, removing all the jobs of task $\tau_j$ arrived before $t_{j+1}$ has no impact on the schedule of the higher-priority jobs (jobs released by $\tau_1, \ldots, \tau_{j-1}$) and the jobs of $\tau_j$ released after or at $t_{j+1}$. 
Moreover, because no task with lower priority than $\tau_j$ executes jobs before $t_{j+1}$ in $\Psi^{j+1}$, removing the jobs released by $\tau_j$ before $t_{j+1}$ does not impact the schedule of the jobs of $\tau_{j+1}, \ldots, \tau_{k}$. Therefore, $t_j$ is set to $t_{j+1}$ and $\Psi^j$ is generated by removing all the jobs of task $\tau_j$ arrived before $t_{j+1}$. The response time of $J_{k}$ in $\Psi^j$ thus remains unchanged in comparison to its response time in $\Psi^{j+1}$. 
\item $J_{j}$ did not complete its execution by $t_{j+1}$. For such a case, $t_{j}$ is set to $r_j$ and hence $\Psi^j$ is built from $\Psi^{j+1}$ by removing all the jobs released by $\tau_j$ before $r_j$. 
\end{itemize}
Note that because by the construction of $\Psi^{j+1}$ and hence $\Psi^j$ there is no job with priority lower than $\tau_j$ available to be executed before $t_{j+1}$, the maximum amount of time during which the processor remains idle within $[t_j, t_{j+1})$ is at most $S_j$ time units.
\item $\tau_j \in {\bf T}_2$, i.e., $S_j \geq C_j$. For such a case, we set $t_{j}$ to $t_{j+1}$. Let $c_j(t_j)$ be the remaining execution time for the job of task $\tau_j$ at time $t_j$. We know that $c_j(t_j)$ is at most $C_j$. Since by the construction of $\Psi^j$, all the jobs of $\tau_j$ released before $t_j$ are removed, the job of task $\tau_j$ arrived at time $r_j$ ($< t_j$) is replaced by a new job released at time $t_j$ with execution time $c_j(t_j)$ and the same priority than $\tau_j$. Clearly, this has no impact on the execution of any job executed after $t_j$ and thus on the response time of $J_k$. The remaining execution time $c_j(t_j)$ of $\tau_j$ at time $t_j$ is called the \emph{residual workload} of task $\tau_j$ for the rest of the proof.
\end{enumerate}
%\end{itemize}
 
The above construction of $\Psi^{k-1}, \Psi^{k-2}, \ldots, \Psi^1$ is repeated until producing $\Psi^1$. The procedures are well-defined. Therefore, it is guaranteed that $\Psi^1$ can be constructed. Note that after each iteration, the number of jobs considered in the schedule have been reduced, yet without affecting the response time of $J_k$. 
%(Note that $J_j$ is defined as the carry-in job of task $\tau_j$ at time $t_k$.) Therefore, the reduced schedule after the above procedure does not change the execution of $J_j$ after time $t_j$ if $\tau_j$ is in ${\bf T}_1$. For a task $\tau_j$ in ${\bf T}_2$, its corresponding carry-in job $J_j$ may be changed, but its execution after $t_j$ remains identical as in the original schedule. 
%Therefore, the resulting schedule above does not change any execution behavior of the (at most) $k-1$ carry-in jobs at time $t_k$.

~

\noindent{\bf Step 2: Analyzing the final reduced schedule $\Psi^1$}

We now analyze the properties of the final schedule $\Psi^1$ in which all the unnecessary jobs have been removed. The proof is based on the fact that for any interval $[t_1, t)$, there is 
\begin{equation}
\label{eq:exec_plus_idle}
\operatorname{idle}(t_1, t) + \operatorname{exec}(t_1, t)  = (t - t_1)
\end{equation}
where $\operatorname{exec}(t_1, t)$ is the amount of time during which the processor executed tasks within $[t_1, t)$, and $\operatorname{idle}(t_1, t)$ is the amount of time during which the processor remained idle within the interval $[t_1, t)$.

Because there is no job released by lower priority tasks than $\tau_k$ in $\Psi^1$, the workload released by $\tau_1, \ldots, \tau_k$ within any interval $[t_1, t)$ is an upper bound on the workload $\operatorname{exec}(t_1, t)$ executed within $[t_1, t)$. From case (b) of Step 1, the total residual workload that must be considered in $\Psi^1$ is upper bounded by $\sum_{\tau_i \in {\bf T}_2} C_i$.
%Consequently, for any time $t$ such that $t_1 < t \leq f_k$, the total amount of idle time and residual workload within $[t_1, t)$ is upper bounded by $\sum_{\tau_i \in {\bf T}_1} S_i + \sum_{\tau_i \in {\bf T}_2} C_i = \sum_{i=1}^{j} b_i$. 
Therefore, considering the fact that no job of $\tau_j$ is released before $t_j$ in $\Psi^1$ ($j=1,2,\ldots,k$), the workload released by the tasks (by treating the residual workload in ${\bf T}_2$ as released workload as well) within any time interval $[t_1, t)$ in schedule $\Psi^1$ such that $t_1 < t \leq f_k$ is upper bounded by 
\begin{align*}
\sum_{i=1}^k \left( c_i(t_i) + \max\{0, \ceiling{\frac{t- t_i}{T_i} } C_i \} \right) & \leq \sum_{\tau_i \in {\bf T}_2} C_i + \sum_{i=1}^k \max\{0, \ceiling{\frac{t- t_i}{T_i} } C_i \},
\end{align*}
leading to 
\begin{equation}
\label{eq:exec_time}
\forall t \mid t_1 \leq t < f_k,\quad \operatorname{exec}(t_1, t) \leq \sum_{\tau_i \in {\bf T}_2} C_i + \sum_{i=1}^k \max\{0, \ceiling{\frac{t- t_i}{T_i} } C_i.
\end{equation}

Furthermore, from case (a) of Step 1, we know that the maximum amount of time during which the processor is idle in $\Psi^1$ within any time interval $[t_1, t)$ such that $t_1 < t \leq t_k$, is upper bounded by $\sum_{\tau_i \in {\bf T}_1} S_i$. That is,
\begin{equation}
\label{eq:idle_time}
\forall t \mid t_1 \leq t < t_k,\quad \operatorname{idle}(t_1, t) \leq \sum_{\tau_i \in {\bf T}_1} S_i.
\end{equation}

Hence, injecting Eq. \eqref{eq:exec_time} and Eq. \eqref{eq:idle_time} into Eq. \eqref{eq:exec_plus_idle}, we get
\[
\forall t \mid t_1 \leq t < t_k,\qquad  \sum_{\tau_i \in {\bf T}_1} S_i + \sum_{\tau_i \in {\bf T}_2} C_i + \sum_{i=1}^{k} \max\{ 0, \ceiling{\frac{t- t_i}{T_i} } C_i\}  \geq t-t_1.
\]
Since $C_k' > 0$ and $\max\{ 0, \ceiling{\frac{t- t_k}{T_k} } C_k'\} = 0$ for any $t$ smaller than $t_k$, it holds that
\[
\forall t \mid t_1 \leq t < t_k,\qquad  \sum_{\tau_i \in {\bf T}_1} S_i + \sum_{\tau_i \in {\bf T}_2} C_i + C_k' + \sum_{i=1}^{k-1} \max\left\{ 0, \ceiling{\frac{t- t_i}{T_i} } C_i\right\}  > t-t_1,
\]
and using the definition of $b_i$
\begin{equation}
\label{eq:eq1_in_proof}
\forall t \mid t_1 \leq t < t_k,\qquad  C_k'+\sum_{i=1}^{k-1} b_i + \sum_{i=1}^{k-1} \max\left\{ 0, \ceiling{\frac{t- t_i}{T_i} } C_i\right\} > t-t_1.
\end{equation}

Furthermore, because $J_k$ is released at time $t_k$ and does not complete its execution before $f_k$, it must hold that
\begin{equation}
\label{eq:eq2_in_proof}
\forall t \mid t_k \leq t < f_k,\qquad  C_k'+\sum_{i=1}^{k-1} b_i + \sum_{i=1}^{k-1} \max\left\{ 0, \ceiling{\frac{t- t_i}{T_i} } C_i\right\} > t-t_1.
\end{equation}

Combining Eq. \eqref{eq:eq1_in_proof} and Eq. \eqref{eq:eq2_in_proof}, we get
\begin{equation}
\label{eq:whole_interval}
\forall t \mid t_1 \leq t < f_k,\qquad  C_k'+\sum_{i=1}^{k-1} b_i + \sum_{i=1}^{k-1} \max\left\{ 0, \ceiling{\frac{t- t_i}{T_i} } C_i\right\} > t-t_1.
\end{equation}

Since $t_i \geq t_1$ for $i=1,2,\ldots,k$, there is 
$$\ceiling{\frac{t- t_i}{T_i} } \leq \ceiling{\frac{t- t_1}{T_i} },$$
thereby leading to
\begin{equation}
\label{eq:with_t1_only}
\forall t \mid t_1 \leq t < f_k,\qquad  C_k'+\sum_{i=1}^{k-1} b_i + \sum_{i=1}^{k-1} \max\left\{ 0, \ceiling{\frac{t- t_1}{T_i} } C_i\right\} > t-t_1.
\end{equation}
By replacing $t-t_1$ with $\theta$, Eq. \eqref{eq:with_t1_only} becomes\footnote{We take $0 < \theta$ instead of $0 \leq \theta$ since $C_k'$ is assumed to be positive.}
\[
\forall \theta \mid 0 < \theta < f_k-t_1, \qquad C_k'+\sum_{i=1}^{k-1} b_i + \sum_{i=1}^{k-1} \ceiling{\frac{\theta}{T_i} } C_i > \theta.
\]
The above inequation implies that the minimum $t$ such that $C_k'+\sum_{i=1}^{k-1} b_i + \sum_{i=1}^{k-1} \ceiling{\frac{t}{T_i} } C_i \leq t$ is larger than or equal to $f_k-t_1$. And because by assumption the worst-case response time of $\tau_k'$ is equal to $f_k-t_k \leq f_k - t_1$ which is obviously smaller than or equal to $R_k'$, it holds that $R_k'$ is a safe upper bound on the worst-case response time of $\tau_k'$.
\end{proof}

For the simplicity of presentation, we assume that $\Psi$ is a schedule of $\tau'$ that generates the worst-case response time of $\tau_k'$ in the proof of Theorem~\ref{theorem:critical}. This can be relaxed to start from an arbitrary job $J_k$ in any fixed-priority schedule by using the same proof flow with similar arguments.

\begin{Corollary}
\label{cor:critical}
 $R_k'$ is a safe upper bound on the worst-case response time of task $\tau_k'$ if $R_k'$ is not larger than $T_k$.
\end{Corollary}
\begin{proof}
Directly follows from Theorem~\ref{theorem:critical}.
\end{proof}

\begin{Corollary}
$R'_k$ is a safe upper bound on the worst-case response time of task $\tau_k$ if $R_k'$ is not larger than $T_k$. 
\end{Corollary}
\begin{proof}
Since, as proven in \cite{Rajkumar_1990,Liu_2014}, the worst-case response time of $\tau_k'$ is always larger than or equal to the worst-case response time of $\tau_k$, this corollary directly follows from Corollary~\ref{cor:critical}. 
\end{proof}

Note that the proof of Theorem~\ref{theorem:critical} does not require to start from the schedule with the worst-case response time  for $\tau_k'$. The analysis still works well by considering any job with any arbitrary fixed-priority schedule.   
To illustrate Step 1 in the above proof, we also provide one concrete example. Consider a task system with the following 4 tasks:
\begin{itemize}
\item $T_1 = 6, C_1 = 1, S_1 = 1$,
\item $T_2 = 10, C_2 = 1, S_2 = 6$,
\item $T_3 = 18, C_3 = 4, S_3 = 1$,
\item $T_4 = 20, C_4 = 5, S_4 = 0$.
\end{itemize}

\begin{figure}[t]
  \centering
  \input{figures/example/example-body.tex}
\caption{An illustrative example of the proof.}
\label{fig:example}  
\end{figure}
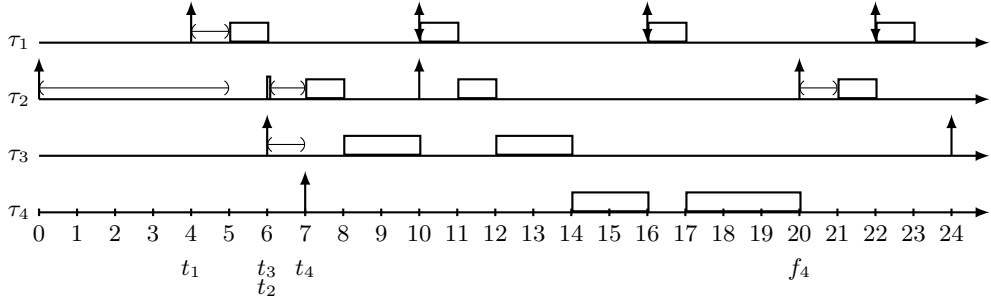

Figure~\ref{fig:example} demonstrates a schedule for the jobs of the
above 4 tasks. We assume that the first job of task $\tau_1$ arrives
at time $4+\epsilon$ with a very small $\epsilon > 0$. The first job
of task $\tau_2$ suspends itself from time $0$ to time $5+\epsilon$,
and is blocked by task $\tau_1$ from time $5+\epsilon$ to time
$6+\epsilon$. After some very short computation with $\epsilon$ amount
of time, the first job of task $\tau_2$ suspends itself again from
time $6+2\epsilon$ to $7$.   In this schedule, $f_k$ is set to $20-\epsilon$.

We define $t_4$ as $7$. Then, we set $t_3$ to $6$. When considering
task $\tau_2$, since it belongs to ${\bf T}_2$, we greedily set $t_2$
to $t_3=6$ and the residual workload $C_2'$ is $1$. Then, $t_1$ is set
to $4+\epsilon$. In the above schedule, the idle time from
$4+\epsilon$ to $20-\epsilon$ is at most $2 = S_1+S_3$. We have to
further consider one job of task $\tau_2$ arrived before time $t_1$
with execution time $C_2$.

%%% Local Variables:
%%% mode: latex
%%% TeX-master: "authorea_build/authorea_paper.tex"
%%% End:

\bibliography{bibliography/biblio}{}

\end{document}

%% file: figures/example/example-body.tex
		\def\ux{0.25cm}\def\uy{0.5cm} 
		\begin{tikzpicture}[y=\uy, font=\sffamily,thick]  
		\tikzset{
			task/.style={fill=#1,  rectangle, text height=.3cm},
			task1a/.style={task=green!30},
			task1b/.style={task=green},
			task2a/.style={task=orange!30},
			task2b/.style={task=orange, minimum width=1mm},
			task3a/.style={task=pink, minimum width=1mm},
			task3b/.style={task=pink!80, minimum width=1mm},
			task4a/.style={task=cyan, minimum width=1mm},
			task4b/.style={task=cyan!50},
			task5/.style={task=blue},
			task6/.style={task=purple},
			task7/.style={minimum height=\ux,draw},
			task8/.style={minimum height=\ux,draw,thick},
			task9/.style={task=gray,minimum height=0.7cm,draw},
		}
		\tikzstyle{jobs}=[ fill=black!50];

		\begin{scope}[shift={(0,0)}]
		
		\draw[<-](2,1.1) -- (2,0);
		\draw[<->](5,1.1) -- (5,0);
		\draw[<->](8,1.1) -- (8,0);
		\draw[<->](11,1.1) -- (11,0);

		\draw[->] (0,0) node[anchor=east] {$\tau_1$}-- coordinate (xaxis) (12.5,0);

		\node[task7, minimum width=2*\ux,
		anchor=south west]at (2.5, 0){};
		\node[task7, minimum width=2*\ux,
		anchor=south west]at (5, 0){};
		\node[task7, minimum width=2*\ux,
		anchor=south west]at (8, 0){};
		\node[task7, minimum width=2*\ux,
		anchor=south west]at (11, 0){};
		\draw[(-), thin] (2, 0.3) -- (2.5, 0.3);

		\end{scope}

		\begin{scope}[shift={(0,-1.5)}]
		%%timeline

		\draw[<-](0,1.1) -- (0,0);
		\draw[<-](5,1.1) -- (5,0);
		\draw[<-](10,1.1) -- (10,0);

		\draw (3,0) -- (3, 0.3cm) -- (3.04, 0.3cm) --
                (3.04, 0);
		anchor=south west]  at ( 3, 0){};
		\node[task7, minimum width=2*\ux,
		anchor=south west] at (3.5, 0){};
		\node[task7, minimum width=2*\ux,
		anchor=south west] at (5.5, 0){};
		\node[task7, minimum width=2*\ux,
		anchor=south west] at (10.5, 0){};

		\draw[->] (0,0)node[anchor=east] {$\tau_2$} -- coordinate (xaxis) (12.5,0);
		\draw[(-), thin] (0, 0.3) -- (2.5, 0.3);
		\draw[(-), thin] (3.04, 0.3) -- (3.5, 0.3);
		\draw[(-), thin] (10, 0.3) -- (10.5, 0.3);

		\end{scope}

		\begin{scope}[shift={(0,-3)}]
		%%timeline
		
		\draw[<-](3,1.1) -- (3,0);
		\draw[<-](12,1.1) -- (12,0);

		\draw[->] (0,0)node[anchor=east] {$\tau_3$} -- coordinate (xaxis) (12.5,0);
		
		\node[task7, minimum width=4*\ux,
		anchor=south west]  at ( 4, 0){};
		\node[task7, minimum width=4*\ux,
		anchor=south west]  at ( 6, 0){};
		
		\draw[(-), thin] (3, 0.3) -- (3.5, 0.3);
		
		\end{scope}

		\begin{scope}[shift={(0,-4.5)}]

		\draw[->] (0,0) node[anchor=east] {$\tau_4$}-- coordinate (xaxis) (12.5,0);
		\draw[<-](3.5,1.1) -- (3.5,0)node[anchor=north] {$$};
				\node[task7, minimum width=4*\ux,anchor=south west]at (7, 0){};
		\node[task7, minimum width=6*\ux,anchor=south west]at (8.5, 0){};
		
		\foreach \x in {0,...,24}{
			\draw[-](.5*\x,0.1) -- (.5*\x,-0.1)
			node[below] {\small $\x$};
			
		}

                \draw (10, -1.5) node {\small $f_4$};
                \draw (3.5, -1.5) node {\small $t_4$};
                \draw (3, -1.5) node {\small $t_3$};
                \draw (3, -2) node {\small $t_2$};
                \draw (2, -1.5) node {\small $t_1$};
		\end{scope}
		
		\end{tikzpicture}  